# Thermoelectric stack sample cooling modification of a commercial atomic force microscopy

A. del Moral, [1,2] J.C. González-Rosillo[T,2] A. Gómez*,[2] T. Puig,[2] and X. Obradors[2]

[1] *Instituto de Microelectrónica de Barcelona, Centro Nacional de Microelectrónica (CSIC), Campus U.A.B., 08193, Bellaterra, Barcelona, Spain*

[2]*Institut de Ciència dels Materials de Barcelona, UAB Campus, 08193 Bellaterra, Spain*

[T] *Currently at Electrochemical Materials Laboratory, Massachusetts Institute of Technology, Bldg. 8-24077 Massachusetts Avenue Cambridge, MA 02139, USA*

* Corresponding autor: agomez@icmab.es

## ABSTRACT

Enabling temperature dependent experiments in Atomic Force Microscopy is of great interest to study materials and surface properties at the nanoscale. By studying Curie temperature of multiferroic materials, temperature based phase transition on crystalline structures or resistive switching phenomena are only a few examples of applications. We present an equipment capable of cooling samples using a thermoelectric cooling stage down to -61.4º C in a 15x15 mm sample plate. The equipment uses a four-unit thermoelectric stack to achieve maximum temperature range, with low electrical and mechanical noise. The equipment is installed into a Keysight 5500LS Atomic Force Microscopy maintaining its compatibility with all Electrical and Mechanical modes of operation. We study the contribution of the liquid cooling pump vibration into the cantilever static deflection noise and the temperature dependence of the cantilever deflection. A $La_{0.7}Sr_{0.3}MnO_{3-y}$ thin film sample is used to demonstrate the performance of the equipment and its usability by analysing the resistive switching phenomena associated with this oxide perovskite.

I. INTRODUCTION

Below room tempearature experiments in Atomic Force Microscopy (AFM) opens new possibilities in studying temperature dependent physical phenomena.[1–7] It has been shown that the combination of high resolution AFM studies and variable temperature experiments can be used to explain transition temperature phenomena, oxygen vacancies mobility, dopant mobility, water adhesion or ferroelectric-paraelectric phase transitions. [6,8–11] Decreasing temperature bellow ambient conditions in Advanced AFM modes, like current sensing contact mode or Piezoresponse Force Microscopy, opens even a wider set of experiments to tackle.[12,13]

The main issues associated to the design of a sample cooler for AFM include reducing the mechanical noise, thermal drift and electronic noise, at the same time. [14] Gas-based coolers have been proposed as a solution to cool samples; however their implementation is extremely complicated due to the high mechanical vibrations.[15] The reduced physical dimensions of the instruments reduces the possibilities of adapting a standard room temperature AFM to be used at low temperatures. Moreover, cooling the complete box of an AFM is not an option, as the specifications of the equipment will not allow temperatures below 5ºC on commercial equipments. A solution for this problem can be found by using a thermoelectric stack to cool samples using only electrical power. Typical commercial equipments available in the market are restricted to a single thermoelectric element, from which low temperature differences are obtained. A summary of the current equipments and their specifications are shown in Table 1, including major AFM manufacturer and third party companies. More importantly, the available instruments does not provide a sufficient . [16–19]

# Cooling & Heating elements available in the market

| Name | Specifications | | | |
|---|---|---|---|---|
| | Min Temp. | Max Temp | Liquid coolant? | Technology |
| Bruker Hysitron | -10 | 100 | N | Single-peltier module |
| Keysight Cooling plate | -5 | 40 | N | Single-peltier module |
| Asylum CoolerHeater | -30 | 120 | Y | Single-peltier module |
| ParkAFM Temperature | -25 | 180 | N | Single-peltier module |
| **This work** | **-61** | **110** | **Y** | **Multi-element peltier module** |

**Table 1:** Comparison between the current equipment available in the market and the equipment proposed in this manuscript.

In this manuscript, we present the necessary modifications of a standard room-temperature AFM to achieve low temperatures down to -61.4º C. The sample plate proposed has an active area of 15 mm x 15 mm, providing low mechanical vibrations and a low electronic noise level. We compare the AFM static deflection noise with and without connecting the equipment to demonstrate the viability of the proposed design. We evaluate the thermal drift phenomena in Contact Mode operation using a commercial AFM probe. The equipment is then tested in the Current Sensing mode to study Resistive Switching (RS) transitions in a 10-12 nm $La_{0.7}Sr_{0.3}MnO_{3-y}$ (LSMO) thin film grown by Chemical Solution Deposition (CSD).

## II. IMPLEMENTATION OF COOLING EQUIPMENT

The implementation of the equipment was done through the scheme showed in **Figure 1**. We added the necessary elements to enable sample cooling to a classic Contact Mode AFM Setup. The equipment design was divided in three different sections: the cooling element, the mechanical setup and the electronic system.

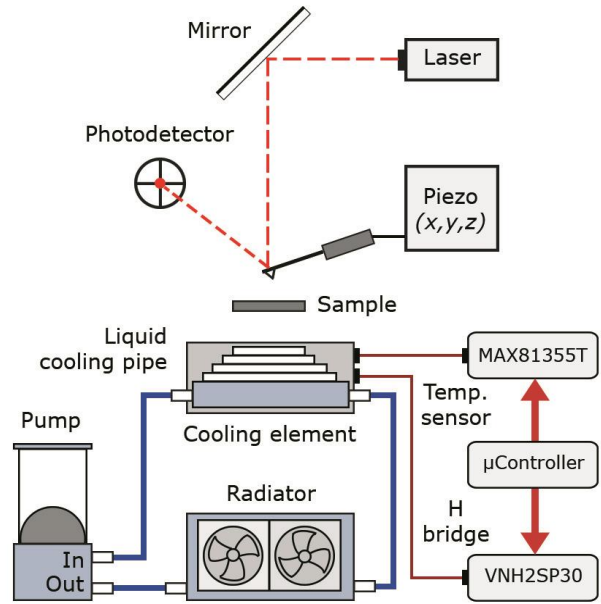

**Figure 1.** Scheme of the modifications implemented in the AFM to enable sample cooling. Blue lines indicate water pipes, while red lines indicate an electrical connection. A four stage thermoelectric cooler is used as cooling element, while refrigerating the hot side of the thermoelectric cooler with a liquid cooling system.

**A. Cooling element**

A thermoelectric-based cooler is chosen as the cooling element for its capability to provide a temperature gradient using only electrical power.[20] The main advantage of this solution relays into avoid using complex gas or liquid nitrogen-based elements. The thermoelectric element is composed of a stack of individual thermoelectric cells pile up in series. The physical X-Y size of each thermoelectric stage decreases from the cooling element to the sample holder as it can be seen in **Figure 1**. With such design the temperature difference between hot and cold plate is maximized. [21,22]

A four element thermoelectric stack is chosen in this equipment to maximize the temperature range. We selected a commercially available thermoelectric energy generator (TEG) stack with reference number TEC4-24603, with 14.6 V input and maximum current of 3 A. According to the data sheet, the maximum temperature difference between hot and cold side of the stack reaches 107 ºC which represents one of the highest values commercially available. The largest refrigeration power of this unit is 6.84 W, which has a maximum power consumption of 45.26 W therefore the efficiency is around 10%, as expected for state of the art thermoelectrics. With the aim of reducing thermal drift, the TEG unit was selected for its ceramic nature and its pyramidal structure, in which the intermediate steps are thermally insulated. The hot side needs to be as close as possible to the ambient temperature. To achieve this goal, a liquid cooling system is used to dissipate the thermal energy from the hot side of the TEG.

**B. Mechanical setup**

The mechanical setup comprises a liquid cooling system and a sample holder. The liquid cooling circuit is composed of a cooling block, a radiator, two fans and an electric pump. Silicone hoses are used because of their flexibility and to avoid vibration transmission from the pump to the sample holder. A water-cooled cooper heat exchanger is stuck to the TEG hot-side face with thermal paste, to minimize thermal losses and restrict the vibration transmission to the block. The block has the water inlet and outlet in one side, which eases the implementation into the AFM. The model "MAN-KD Full Copper HZ Top Version" block was selected while "Artic Silver Ceramique 2" was used as thermal compound. Any other model or brand that is suitable can be used as a water cooling block to dissipate the heat from the thermoelectric element. For the water cooling pump, a key requirement is to keep both noise and vibration levels as low as possible. The DC pump MS-500 is selected because of it is low noise, with a maximum noise level of 16 dB and maximum static pressure of 3 meters. The pump is mounted in the equipment box using special rubber bands to avoid vibration transmission while it is voltage input is compatible with the power supply used. The circuit includes a double copper radiator that dissipates water heat using two 12 V, 12 cm of diameter, fans. The copper radiator selected is a "Black Ice Stealth GT240", however any commercial double cooper water radiator can be employed. Galvanic corrosion has been prevented by using copper in both block and radiator, besides the fact that copper heat sinks offer an optimal heat transmission. [23]

**C. Electronics**

The TEG used, TEC4-24603, has its maximum efficiency at 13.4 V. We selected a DC voltage source of this exact same voltage; model number FDPS-360W that also supplies current to the pump and the fans. Both pump and fans are designed for 12 Volts operation, but despite the slightly higher voltage every component can be feed with the same power supply. A Pulse-Width Modulation (PWM) signal is used to control the power supplied to the thermoelectric cooler, through an electronic current controller, a H bridge, allowing both cooling and heating. The temperature on top of the sample plate is measured with a Thin Film Platinum Resistance PT1000J sensor. We selected this sensor because of its physical size of 2.3 x 2 mm and its low temperature error of 0.3 ºC. A SAM3X8E microcontroller is programmed with a Proportional-Integral-Derivative (PID) controller to set the desired temperature, however commercially available PID controllers can be used to set the temperature through a Solid State Relay (SSR).

### III. Characterization of the equipment

We tested the performance of our equipment through different tests. Before mounting the cooling element in the AFM box, we completed a performance test of the lowest achievable temperature at the cold side of the thermoelectric element for different voltages supplied. The results are plotted in **Figure 2a**.

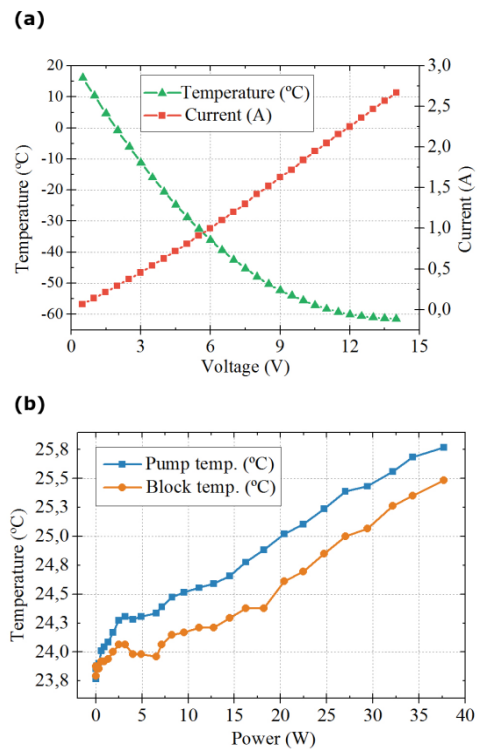

**Figure 2a**. Temperature and Current versus supply Voltage for the cooling equipment. A resistor like behaviour can be seen, with a maximum current of 2.7 A. The lowest temperature point is 13.4 V, increasing the voltage from this point does not lower the temperature. **Figure 2b**. Liquid temperature (blue) and thermoelectric hot-side temperature (orange) versus power dissipated at the thermoelectric cooler. It can be seen that the hot-side temperature increases less than 2 ºC at maximum power.
.

The data is acquired by increasing the supply voltage of the thermoelectric element in 0.5V steps, while waiting 3 minutes for temperature stabilization before each measurement. From the data, it can be seen that the minimum achievable temperature is -61.4 ºC at a voltage of 13.4 V. From 13 V to 14 V the thermoelectric effect cannot compensate Joule heating and, as a consequence, the cold-side temperature reaches its minimum.[24] Current measurement is investigated at each applied bias to characterize the power consumption of the element, finding the maximum current drained of 2.67 A while the element has a resistor like behaviour. A key factor to reach low temperatures is the efficiency into dissipating the heat out of the hot side of the TEG. To test the design of the liquid cooling system, we measure the temperature of the water inside the liquid cooling system and the temperature at the hot side of the thermoelectric. We found that, at the maximum power dissipation of 37.4 W, the temperature of the water is increased by merely 2.0 ºC, from ambient temperature, while the external temperature of the water cooling block increased 1.7 ºC above room temperature. The increased temperature versus power dissipation is shown in **Figure 2b**.

Another key factor for the usability of the cooler inside the AFM, is the vibration noise caused by the pump. We selected a low noise, low vibration pump, with a low 16db noise level. In spite of this low noise level, the vibration transmitted through the silicon tubes causes a high frequency noise induced by mechanical vibration located in the sample holder. The vibration was high enough to avoid the acquisition of topography images in dynamic mode. Thus, an anti-vibration system was designed, employing anti-vibration foam elements to isolate vibration transmission from the silicon tubes to the AFM box. To test the noise improvement, we recorded the raw deflection movement of the cantilever in contact mode, with the pump turned off and the pump turned on, with and without anti-vibration system. The results are plotted in **Figure 3**, where the deflection of the cantilever vs time was recorded.

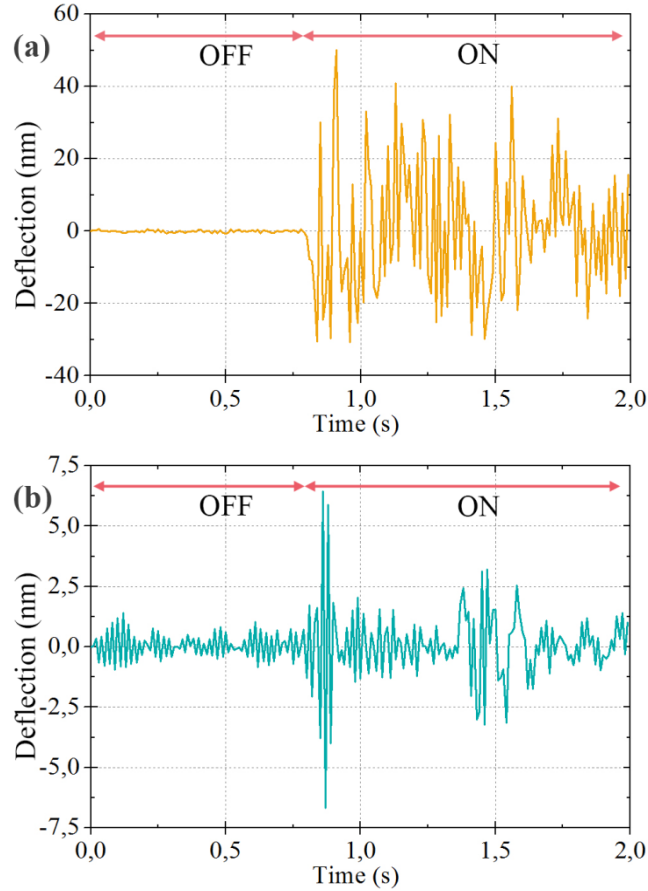

**Figure 3a** Cantilever Deflection versus time without the anti-vibration system installed. **Figure 3b** Noise versus time with the anti-vibration system installed. The arrows denote where the pump was turned on during the data collecting period, note the different Y-axis scales.

**Figure 3a** shows the effect of turning the pump on with the absence of anti-vibration system, while **Figure 3b** shows the improvement after anti-vibration elements were installed. From the data, we found that the maximum static deflection noise is 0.6 nm without the pump and 1.6 nm with the pump connected. Without the anti-vibration system, the vibration increases dramatically to 17.7 nm with the pump connected.

## IV. PERFORMANCE OF THE SYSTEM

To test the equipment, we used a $La_{0.7}Sr_{0.3}MnO_{3-y}$ thin film sample of 10-12 nm thickness, which shows Resistive Switching (RS) characteristics. It is known as RS the ability of certain materials, mainly oxides, to change its resistance state in a reversible way under the application of a high electric field[22-23]. LSMO thin films present bipolar RS effect, and its IV characteristics can be measured with a CS-AFM as the nanometric size of the tip is able to generate locally very intense electric fields. In these metallic complex oxides systems the RS effect can be attributed to oxygen vacancies motion and therefore, an electrical study below room temperaturecan provide more evidence into this issue. Topography images were obtained using App Nano FORT tip, working in the dynamic constant amplitude mode. We recorded topography images at ambient temperature, with and without the pump connected, to see the influence of the vibration induced by the pump. We then obtained an image at a minimum temperature of -52.8º C, with the pump connected. The results are showed in Figure 4. We conclude that the topography can be revealed with the pump connected or disconnected in the whole temperature range.

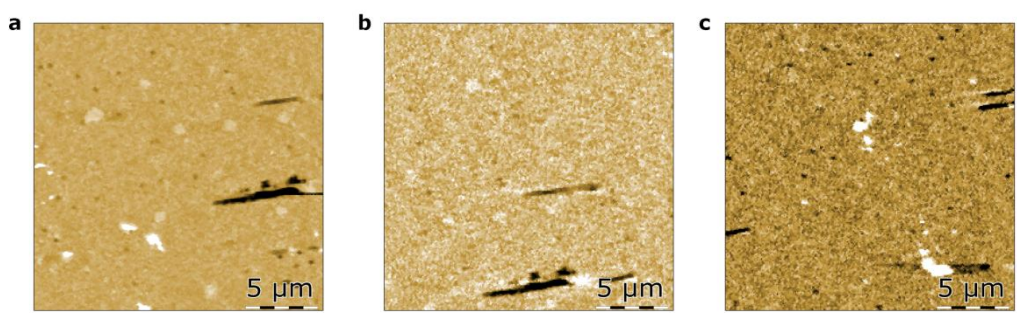

**Figure 4a** 20 μm x 20 μm Topography image obtained at ambient temperature with pump turned off. **Figure 4b** 20 μm x 20 μm Topography image obtained at ambient temperature with pump turned on, **Figure 4c** 20 x 20 μm topography image obtained at -52.8 ºC with the pump turned on. Z scale bar is 80 nm.

An AFM advanced mode called Current AFM is used to investigate the use of the new equipment in other operation circumstances. In this case, a conductive, doped-diamond tip, model DDESP-V2, is used to perform current measurements of the sample. Both doped-diamond and full platinum tips can be used for taking the measurements[25]. In our setup, the sample is biased with an applied DC voltage while the tip is grounded. We performed current versus voltage curves to study the RS phenomena of the sample, at different temperatures. The results are plotted in **Figure 5**.

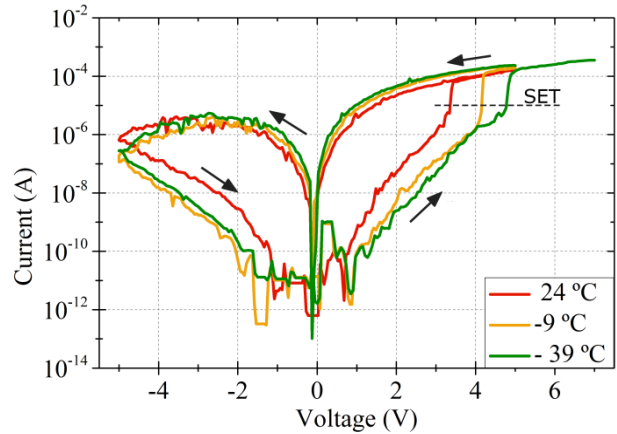

**Figure 5**. Current vs Voltage curves at different temperatures, Y-axis is in absolute logarithmic scale. The sweep starts from 5 VDC towards negative bias in the low resistance state (LRS). At negative bias, a transition from LRS to high resistance state (HRS) can be observed. The cycle continues from negative values towards positive values, until the initial LRS state is recovered. The cycle was done at different temperatures, to see the evolution of the SET voltages.

The voltage sweep starts from 5 VDC towards negative bias, then it goes from negative voltage back to positive 5 VDC. For every selected temperature a full RS transition cycle can be observed. The RESET process consisting of a transition from the high resistance state (HRS) to a low resistance state (LRS) occurs at positive bias. The reverse transition, known as SET process, recovers the initial LRS state at negative values. When the temperature is decreased, a shift towards higher values of the SET voltage is observed. For the -39ºC sweep, it was even necessary to apply a higher voltage (+7V instead of +5V) to observe the full transition from HRS back to LRS. We can quantify the evolution of the SET voltage for a given threshold current of $10^{-5}$ A The set voltage is 3.4 V for 24ºC, 4.2 V for -9ºC and 4.8 V for -39ºC. Careful analysis of this kind of measurements can provide more insight into the physical mechanism underneath the RS mechanism in complex oxides systems.

Finally, we performed a thermal drift test to measure the vertical expansion and contraction of the equipment versus temperature. The test is done with the tip engaged to the surface of the sample. A temperature of -52.8 ºC is set and then the cooler was turned off. With the cooler turned off, the temperature of the sample plate will increase up to room temperature in 150 seconds. We record how the Topography channel changes versus time, from -52.8ºC until room temperature is reached, results are plotted in Figure 6.

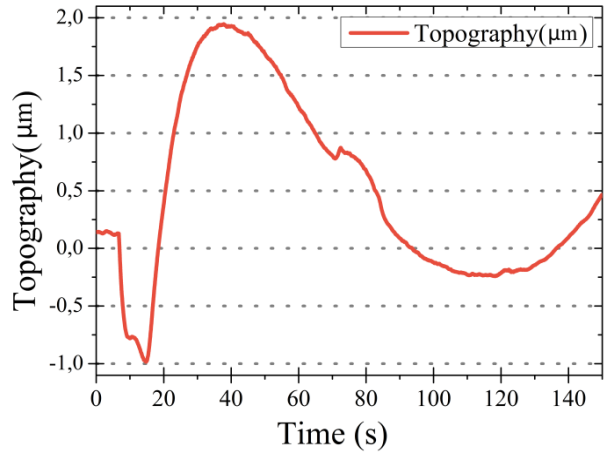

**Figure 6**. Topography (μm) vs Time (s), curve obtained from -52 8 ºC to ambient temperature. Maximum thermal expansion is found to be less than 3 μm.

The thermal expansion in Z direction of the equipment is less than 3 μm, giving us the possibility to heat and cool samples without withdrawing the AFM probe from the sample, as the Z axis piezo range is 9 μm.

## V. CONCLUSIONS

A sample cooling stage system has been implemented and tested as an upgrade to commercially available Atomic Force Microscopy (AFM) equipment. A four-stage thermoelectric cooler is used as sample cooler with a water cooling system to complete the setup. A characterization of the performance of the cooling equipment has been reported and we have demonstrated that we can reach temperatures down to -61 ºC, with a maximum power consumption of 37.4 W and a maximum thermal expansion of less than 4 um. The viability of the setup is demonstrated both in Dynamic mode, to obtain topographic images, and Current Sensing mode, to electrically characterize samples. The Resistive Switching behaviour of a $La_{0.7}Sr_{0.3}MnO_{3-y}$ thin film sample versus temperature has been provided. Data show an increase of the voltage required to recover the low resistance state with decreasing temperature. These results will lead to a better understanding of the microscopic mechanism underlying the resistive switching behaviour of complex systems.

### VII. Acknowledgments

This research was supported by Consolider NANOSELECT (CSD 2007-00041). ICMAB acknowledges financial support from the Spanish Ministry of Economy and Competitiveness, through the "Severo Ochoa" Programme for Centres of Excellence in R&D (SEV- 2015-0496). The authors thank ICMAB Scientific and Technical Services.